# A comprehensive safety engineering approach for software-intensive systems based on STPA

Asim Abdulkhaleq[a,*], Stefan Wagner[a], Nancy Leveson[b]

[a]*Institute of Software Technology, University of Stuttgart, Stuttgart, 70174, Germany*
[b]*Massachusetts Institute of Technology, 77 Massachusetts Avenue, Cambridge, MA 02139-4307, United States*

**Abstract**

Formal verification and testing are complementary approaches which are used in the development process to verify the functional correctness of software. However, the correctness of software cannot ensure the safe operation of safety-critical software systems. The software must be verified against its safety requirements which are identified by safety analysis, to ensure that potential hazardous causes cannot occur. The complexity of software makes defining appropriate software safety requirements with traditional safety analysis techniques difficult. STPA (Systems-Theoretic Processes Analysis) is a unique safety analysis approach that has been developed to identify system hazards, including the software-related hazards. This paper presents a comprehensive safety engineering approach based on STPA, including software testing and model checking approaches for the purpose of developing safe software. The proposed approach can be embedded within a defined software engineering process or applied to existing software systems, allow software and safety engineers integrate the analysis of software risks with their verification. The application of the proposed approach is illustrated with an automotive software controller.

## 1. Introduction

Given the rapid innovations in software and technology, many complex systems are becoming software intensive. Software-intensive systems are systems in which software interacts with other software, systems, devices, sensors and with people [1]. Software safety becomes a critical aspect in the development of modern systems. The most widely practiced safety analysis approaches. Fault Tree Analysis (FTA) [2], Failure Mode and Effect Criticality Analysis (FMECA) [3] and Hazard and Operability Analysis (HAZOP) [4] were all developed over 50 year ago, before computers were common in engineered systems. As a result, these safety analysis methods do not completely ensure safety in complex systems. In such systems, the accidents result from a number of different factors such as software requirements flaws, hardware errors, human mistakes, environmental influences and especially the interaction among them. A new trend is to advance safety analysis techniques using system and control theory rather than reliability theory. STAMP (Systems-Theoretic Accident Model and Processes) [5] is a modern approach to safety engineering that promises to overcome the problems of traditional safety analysis techniques. STPA is designed for safety analysis in the system development and operating stages; the goal is to identify hazards existing in the system and provide safety constraints to mitigate those hazards.

* Corresponding author. Tel +49 711 685-88455; fax: +49 711 685-88380.
E-mail address: Asim.Abdulkhaleq@informatik.uni-stuttgart.de

System safety engineering is an engineering discipline to identify system hazards and prevent systems from getting into unsafe (hazardous) states. Many safety standards have been introduced, such as ISO 26262 [6] and MIL-STD-882 [7]. These standards define a generic engineering process to develop safe systems, however, they do not mandate the use of specific techniques or methods to be used during the system development process. They also neither mandate any safety analysis technique nor give concrete guidance on how to identify and classify hazard scenarios, i.e., the causes of hazards.

***Problem Statement***: A software safety process involves the following steps:(1) software safety requirements must be identified at the system level and (2) safety engineers must ensure that the software meets the safety requirements and will not transition into a hazardous state. STPA has been developed to derive detailed safety requirements for complex systems. However, the subject of STPA is the system and not only the software. STPA treats the software components like other physical components of the overall system. Moreover, STPA has not yet been placed into the software development process of safety-critical systems and the current software engineering methods do not explicitly incorporate STPA safety activities. STPA safety analysis is often handled separately by safety engineers, while software developers are usually not familiar with system safety analysis processes. Therefore, there is a gap between software and safety engineering processes. Traditionally, testing and formal verification approaches are used in the development process to assert the functional correctness of software, which usually is not enough ensure the safe operation of software.

***Research Objectives:*** The main objective of this research is to fill the aforementioned gap by integrating STPA safety activities in a software engineering process, thus offering seamless safety analysis and verification. This will help engineers derive the software safety requirements, verify them, generate safety-based test cases, execute them to recognize the associated software risks, and reduce them to a low level.

***Contribution:*** We propose a safety engineering approach to derive software safety requirements at the system level and verify them at the design and implementation levels. First, we derive the STPA software safety requirements at the system level. Then, we formalize these STPA software safety requirements in a formal specification. To verify the software against the STPA results by using formal verification approaches (e.g. model checker), we build a safe behavior model of a software controller based on system requirement specifications and the STPA results. We use this model to generate safety-based test cases based on the software safety requirements derived by the STPA.

## 2. Background

*2.1. Software safety & STPA*

Software is an integral and increasingly complex part of modern safety critical systems. Therefore, it is essential to analyze software safety in a system context to get a comprehensive understanding of the roles of software and to identify the software-related risks that can cause hazards in the system. Leveson [8] noted that software by itself is not hazardous and cannot directly cause damage to human life or environment; it can only contribute to hazards in a system context. Software can create hazardous system states through erroneous control of the system or by misleading the system operators when taking actions [5]. Software has no random failures and it does not wear out like hardware components. Flaws in software are systematic failures which stem from flawed requirements, design errors or implementation flaws [8, 9]. System hazards related to software are caused by software flaws, software specification errors and uncontrolled interactions among different components forming the system, rather than failures of single components. Ensuring the safe operation of systems requires that the potential risks associated with increased reliance on software be well understood, so that they can be adequately controlled. Therefore, to develop safe software, we first need to identify and analyze software-related hazards and the corresponding software safety requirements at the system level. To assure that these software-related hazardous causes cannot occur in a system, safety verification and testing activities include a demonstration of whether the software design and implementation meet those software safety requirements.

STPA (Systems-Theoretic Processes Analysis) is a safety analysis approach built based on systems theory rather than reliability theory. It can identify system safety-related constraints and potential causes for their violation that are necessary to ensure system safety. STPA can be iterated until the design is set and all hazards are identified and controlled. Unlike traditional safety analysis techniques, STPA provides guidance and a systematic process to identify the potential for inadequate control of the system that could lead to a hazardous state resulting from inadequate control or enforcement of the safety constraints.

### 2.2. Software verification

Software verification is a process to check whether software fully satisfies all the expected requirements. Formal verification and testing are two fundamental approaches to software verification. Formal verification techniques are used to prove the correctness of software and check whether the software satisfies its requirements. Three types of formal verification exist: model checking, theorem proving and deductive methods [10]. Model checking is a well-established formal verification technique to verify whether embedded software for safety-critical systems meets the requirements through exhaustive exploration of the state space of the software. The model checking process involves the target software to be formally modelled in the input language of a model checker and specifications / properties to be formalized in a temporal logic such as Linear Temporal Logic (LTL) [11]. The model checking tool will perform an exhaustive exploration to verify whether the software holds a given property. In case that the software does not hold a given property, a model checker will produce a counterexample that identifies a path where the software violates the given property. There exist two ways to extract the input model of required by software model checker tools: (1) at the design level a verification model can be constructed from the state machine diagram (e.g. SMV model [12]) and (2) at the implementation level a verification model can be extracted directly from software code (e.g. SPIN model [13]).

Software testing is the process of executing software with the intent of finding defects [14]. The main goal of testing is to provide confidence in the correctness of software. Several software testing approaches have been developed to evaluate the quality of software. They do not, however, directly address safety concerns. The complexity of software makes testing a challenging process because usually impossible to test all possible execution paths of software. Model-based testing [15] is a test technique for generating test cases based on the model of the system under test. Model-based testing can be performed in three major steps: modelling the system under test, test case generating and test execution.

### 3. Related work

A new extended approach to STPA was introduced by Thomas [16], whose approach aims to identify unsafe control actions in STPA Step 1 based on the combination of process model variables of each controller in the control structure diagram. Some control actions in the system can only be hazardous in a certain context. Therefore, the process model variables should be assembled to define a context and analyzed based on their context to check if this combination could lead to a hazard or not. Based on that, to make the identified process models more systematic, we integrated the explicit modelling of state machines into the process models of the control structures in STPA [17]. This way we enabled additional state-based analyses and aided the identification of the unsafe control actions of systems in STPA

Step 1 based on the combinations of critical states of the system. We used the idea of the context tables to generate the test cases for each software safety requirement by modelling the STPA results in the safe behavior model of the software controller.

The use of model checking to generate test cases was introduced by Ammann and Black [18]. Amman and Black combined mutation analysis with model-checking based test case generation to produce mutation test cases from the specification. They used SMV model checking to generate test sequences. Based on that, we used the model checking to generate the counterexamples for each software safety requirement to generate safety-based test cases. Utting and Legeard [15] developed an open tool support called ModelJunit, which we used. ModelJUnit is an open source tool support for model-based testing based on the Java library that extends JUnit. ModelJUnit accepts simple Finite State machine (FSM) models or Extended Finite State Machine (EFSM) models as Java classes that encode the states, transactions and guards (constraints) as well., then it generates tests from those models and measures various model coverage metrics. We used ModelJUnit to generate the test case from the safe behavior model due to two main reasons: (1) The safe behavior model can be easily converted into a ModelJUnit Java class and (2) it provides a simple way to generate the test input data.

Our earlier work [19] aimed at providing a software safety verification methodology based on STPA safety analysis to verify the STPA software safety requirements based on the state machine model at the design level. We applied STPA to vehicle cruise control software to identify the software safety requirements at the system level. We used NuSMV to verify these requirements based on an SMV model constructed manually from the software specification. The approach is effective in identifying software safety requirements and verifying them based on an abstract model, but we could not ensure that the SMV model matches exactly the software implementation, so the SMV was converted based on the state machine model of system at the design level. Recently, we investigated the possibility of verifying the software safety requirements based on the model extracted directly from the source code of software by integrating STPA with a software model checker [20]. We have also developed an extensible STAMP platform [21] as open tool support for safety engineering designed specifically to serve the widespread adoption and use of STPA in different areas, which can be easily extended to include different requirements and to support the proposed approach.

**4. The proposed approach**

To develop new safe software for safety-critical systems, a new, more systematic software and safety engineering process is required. For this purpose, we propose a comprehensive safety engineering process based on STPA that combines the safety-related activities to derive the software safety requirements at the system level and verification and testing activities at the design and code levels to formally verify and test whether the software meets these requirements. Fig. 1. An overview of a comprehensive safety engineering approach.

The proposed approach uses STPA to identify the safety requirements of the system and the uses formal verification to verify the functional requirements on a model of the software. Fig. 1 shows an overview of the proposed approach, which is divided into three activities, (1) deriving software safety requirements at the system level; (2) constructing the safe behavior model of the software controller; (3) software safety verification performed with two complementary activities: (3.1) formally verifying software design and implementation against its safety requirements at the design and code levels, and generating safety-based test cases, and (3.2) testing the generated safety-based test cases using the safe behavior model of the software controller.

The proposed approach can be applied during the development process of new software or to existing software. The initial input of safety engineering is the system specification and requirements. Based on these specifications, the safety analyst will perform the STPA safety analysis, which is the starting point of the safety engineering process. Before starting the STPA safety analysis process, the safety analyst will investigate the functional system requirements, which include the software requirements and the system specification documents, gather the fundamentals of analysis, and guide the analysis process in the next steps. In the following sections we describe in more detail the four major activities:

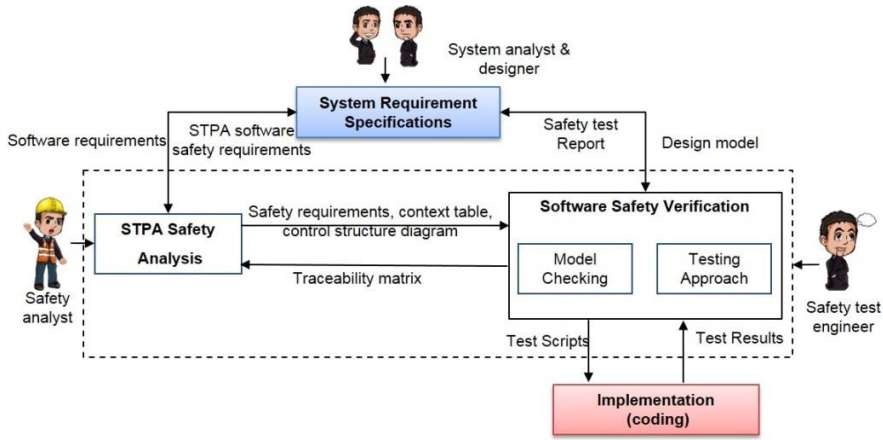

Fig. 1. An overview of a comprehensive safety engineering approach.

### 4.1. Step1: Deriving the software safety requirements at the system level by STPA

Based on the basic procedure of STPA and the extended STPA approach [16] for identifying unsafe control actions based on the combinations of process model variables, we developed an algorithm to explore the software safety requirements at the system level [20]. The algorithm requires that the safety analyst construct the safety control structure diagram of the system from the system specification. The safety control structure is a high-level abstraction diagram that includes the main subsystems that interact with software. The software is the controller in the control structure diagram. The safety analyst can derive the software safety requirements for each software controller in the control structure diagram by performing the following steps:

1. Identify all software safety-critical Control Actions (CA) that can lead to one or more of the defined hazards (HA).
2. Evaluate each CA against four general types of hazardous behaviors to identify the Unsafe Control Actions (UCAs): (1) a control action required for safety is not provided, (2) an unsafe action is provided, (3) a potentially safe control action is provided too early, too late or out of sequence, and (4) a safe control action is stopped too soon or applied too long. Then the identified UCAs are translated into informal textual Software Safety Requirements (SSR).
3. Identify the process model and the variables that affect the safety of the control actions and include them in the software controller in the control structure diagram to document how each UCA could occur. The process model contains three types of variables (states): Internal variables (I) of the software controller, Interaction interface variables (IN), which receive data/command of the environmental components, and Environmental variables (E) of other components in the system interacting with the software controller.
4. Identify the combinations of the process model variables for each unsafe control action. Each combination should be evaluated within two contexts ($C_i$=Providing CA or Not Providing CA) to determine whether the control action is hazardous in that context. A CA could be considered as hazardous in context C if only a combination of process variables related to CA lead to a system-level hazard $H \in HA$. Then use a t-way combinatorial testing algorithm [22] to identify a minimal combination of process model variables for large and complex software.
5. Identify the potentially unsafe scenarios leading to each unsafe software control action and translate each unsafe scenario from the context table into the corresponding software safety constraint by using Boolean operators AND and OR for each context.

The output of this step is a safety control structure diagram with a process model, a list of software safety requirements, and a set of unsafe software scenarios.

*4.2. Step 2: Constructing the safe behavior model of the software controller*

To verify the software design or implementation against the results of STPA and generate the corresponding safety-based test cases, each software controller's behavior must be modelled in a suitable behavior model and constrained by the STPA safety requirements. For this purpose, the system designer and safety analyst should work together to construct a safe behavior model of each software controller based on the system specification and the safety control structure diagram that contains the process model of the software controller. The safe behavior model should be labeled with software safety requirements as derived by STPA. A safe behavior model explores the safety-related behavior of the software controller based on the results of STPA. It includes only the process model variables that affect the safety of control actions of the software controller, their relationship, and the ways in which the system can migrate from one state to another. We used the UML statechart notation to visualize the safe behavior model, which is being used extensively during software development to capture the behavior aspects of the software rather than its physical or functional aspects. The software engineers should map the informal textual STPA software safety requirements into design and implementation specifications (e.g. name of variables and critical functions); this will reduce the time and effort required during the software safety verification step to solve conflicts between the names of safety-critical variables in the LTL formulae and software variables that are used in the software code.

*4.3. Step 3: Software safety verification*

This step aims at verifying the STPA-generated software safety requirements based on the verification model, which is constructed manually from the safe behavior model at the design level or extracted directly from the source code at the implementation level. It aims also at generating safety-based test cases to check whether the software can violate its STPA safety requirements. The input of this step is the safe behavior model and the STPA results. The safety test engineer should perform the verification activities in this step with two complementary tasks: (1) formal verification and (2) testing.

*4.3.1. Formal software safety verification*

During this step, we verify whether the safe behavior model of software satisfies the software safety requirements that we derived by STPA and specified in the LTL, and generate the safety-based test cases. The safety test engineer should perform the software safety formal verification through the following four sub-tasks:
1. Formalize the software safety requirements: Once the safety analyst has identified the unsafe software scenarios and the corresponding software safety requirements by STPA, the safety test engineer has to formalize the corresponding software safety requirements of unsafe scenarios into a formal specification such as the LTL or CTL to be able to verify them during the verification phase;
2. Extract the input verification model of the model checker. The verification model can be constructed in two ways: (a) manually construct the verification model from the safe behavior model into the SMV model at the design level. The verification model will be verified by using the NuSMV model checker, or (b) Extract the verification model directly from the software code by using Modex [23] for ANSI-C Code at the implementation level. The verification model will be verified by using the SPIN model checker; and
3. Verify the safe behavior model against each STPA software safety requirement specified in the LTL/CTL to ensure that the safe behavior model of the software controller includes the STPA safety requirements.

*4.3.2. Generating safety-based Test Cases*

Model-based testing is a process to generate automatically test cases based on a given model of the system under test. The safe behavior model will be used as input to the model-based testing process to generate safety-based test cases. The safety test engineer should perform the model-based testing process through the following four sub-tasks:

(1) Convert the safe behavior model of the software controller into the input language of the selected model-based testing tool; (2) Select a suitable test coverage criteria (i.e. all transactions coverage, all state coverage, or action coverage) to ensure that every transition in the safe behavior model is considered; (3) Generate the model-based test data for the critical variables of the system under testing; and (4) Use a model-based testing tool to perform the traversal on the safe behavior model. During traversal, collect the actions, conditional expressions (i.e. guards) and test input and output data on each transition to generate safety-based test cases.

The output of this step is the test cases to be grouped into test suites. A test suite comprises a group of relevant test cases. The model checking is used here to verify the safe behavior model behavior against the results of STPA to ensure that the safe behavior model includes and satisfies the STPA results. It is also used to generate specific test cases for each software safety requirement. In addition, the safety test engineer will use the model-based testing approach to generate automatically the test cases based on the safe behavior model of the software controller. The main problem with generating test cases with model-based testing is the redundancy of test cases for each safety requirement. Therefore, the safety test engineer should remove manually or automatically the duplicated test cases of each test suite before generating the test scripts.

*4.3.3. Generating and executing test-scripts*

This step aims at generating the executable test scripts of each abstract test suite and executing automatically or manually each test script on the system under test to check whether the software implementation fulfils its software safety requirements. Each test script includes a set of instructions that will be performed on the system under test, the mapping of the test cases to function calls of software code, and test input data of each variable to test that the software behaves as expected. After execution, the test results are analyzed to determine the output of each test execution. To verify how each test script covers the software safety requirements derived by STPA, the safety test engineer should construct manually or automatically the traceability matrix between the test cases and the STPA-generated safety requirements.

The output of the proposed approach is a safety verification report that shows the results of software safety verification activities of each software safety requirement, the coverage measure, the results of both formal verification and model-based testing and the results of executing the relevant test scripts. The system engineers will use the safety verification report to modify the system design or implementation to ensure the safe operation of the software controller.

## 5. Illustrative Example: A Software Controller of an Adaptive Cruise Control System

In this section, we illustrate the proposed approach with an example of the software controller of an Adaptive Cruise Control (ACC) system. In our previous work [19, 20], we described steps 1–4 in more detail. Due to length limitations, we use the same results of step 1 from our previous work to demonstrate briefly how to use the STPA results to construct the safe behavior model of the software controller of the ACC, and how to generate the safety-based test cases. The software controller of an adaptive cruise control system monitors the front road with the use of a radar system attached to the front of the vehicle and keeps a safe distance between the ACC vehicle and a vehicle in the ACC vehicle's path. The software controller interacts with four environmental components, (driver, radar, brake and engine control systems) along with vehicle speed and brake sensors. The accident that the ACC software controller can lead or contribute to is: **ACC vehicle collides with a vehicle in front while ACC is active**. The system-level hazards that can lead to this accident are: $H_1$ = *ACC software controller does not keep a safe distance from a nearby vehicle* and $H_2$ = *Unintended acceleration when the ahead vehicle is too close*.

Table 1 shows examples of the software safety requirements derived from step 1. The ACC software issues two control actions: *accelerating signal* and *brake decelerating signal*. The process model variables of the software that affects the safety of these control actions are: environmental variables (E) (*driver input*), Interaction interface variables (IN) (*radar data, brake status and vehicle speed),* and internal variables (*ACC Mode* which has five values: *on, off, standby, cruisespeed, and followdistance*). The *cruisespeed* variable is a state variable in which there is no forward vehicle in the lane, and the current speed of the vehicle is equal to the desired speed set by the driver. It has three substate values: *cruise* in which the current speed is equal to the desired speed, *accelerate* in which the current speed is less than the desired speed, reduce speed state in which the current speed is greater than the desired speed, and *unknown* where the current speed is not known, for example, an update has not been received for a long time. The *followdistance* variable is a state variable in which a forward vehicle is getting closer in the lane. It has three sub-states: *follow* in which the distance between the ACC vehicle and the target vehicle is being controlled, *decelerate* in which the distance between the vehicles is less or equal than the safe distance, and *resume* in which the distance between the two vehicles is greater than the safe distance.

Table 1. Examples of the corresponding software safety requirements

| ID | Corresponding Software Safety Requirements |
|---|---|
| SSR1.1 | The ACC software controller should provide an acceleration signal when the target vehicle is no longer in the lane |
| SSR1.2 | The ACC software controller decelerates the speed when the distance to the target vehicle is too close. |
| SSR1.3 | The ACC software controller should not provide the acceleration signal speed when a safe distance is reached |
| SSR1.4 | The ACC software controller should not increase the speed beyond the value of the desired speed set by the driver |

Table 2 shows the examples of the context table [16] when providing the control action: *accelerate signal*. The process model variables that have an effect on the safety of providing accelerate signal are: *followdistance* (3 values), *cruisespeed* (3 values), *brake status* (2 values) and *ACC Mode* (4 values). The process model includes *3x3x2x4=72* combinations. To reduce the number of combinations, we first applied a pairwise algorithm that ensures that each two values (pairs) in the context table will appear at least once. Based on that, we reduced the number of combinations between the relevant process model variables from 72 to 12. Next, we applied rules to the context table such as a rule of ACC cancelling: *ACC will cancel only if the button OFF is pressed or the brake pedal is applied*. Based on this rule, we omitted all the rows that contain ACC mode is OFF and brake status is *applied*. The final number of the process model variable combinations was reduced to 6. Each row in the context table was evaluated to determine whether the control action is hazardous in that context. For example, the context of providing an acceleration signal when the current distance is less than or equal to the safe distance, and the current speed is greater than the desired speed and the brake is not pressed is hazardous (i.e. it leads to the hazard H2). Based on the evaluation of each row in the context table, the software safety requirements are refined. For example, the software safety requirement (SSR 1.4) is refined as the ACC software controller should not provide accelerating signal more than the desired speed while ACC is in cruise mode and brake pedal is not pressed.

Table 2. Examples of the context table of providing the control actions

| Control Actions | Process Model Variables | | | | Hazardous? |
|---|---|---|---|---|---|
| | Distance | Speed | Brake | ACC Mode | |
| Accelerate Signal provided | Distance < safe distance | Speed == desired speed | applied | Cruise | No |
| | Distance < safe distance | Speed > desired speed | Not applied | Cruise | Yes (H2), SSR3-4 |
| | Distance < safe distance | Speed > desired speed | Not applied | follow | Yes (H1), SSR1 |

Once we have identified the software safety requirements, the process model and the unsafe scenarios of each control action using step 1, the safe behavior model can be constructed based on the process model. The safe behavior model of the ACC software controller shows the relations between the process model variables (identified by step1) and labeled with software safety requirements. Each transition in the safe behavior model is labeled with the syntax: *event [safety constraint]/control action*. The *event* is a trigger of the transition and the *safety constraint* is a Boolean condition that must be true to transit to the next state. The *control action* describes the effect of the transition, such as how the state variables are updated and what events are generated. For example, the transition $t_6$ can be written:

*controlSpeed(currentspeed) [currentSpeed < desiredSpeed && distance > safeDistance && ACCMode ==cruise && Brakestatus == Notapplied ]/ accelerateSpeed(currentspeed).*

The transition t$_6$ constrains the provision of the *accelerate* control action under the safety constraint derived by step 1 (Table 2). To formally verify the software safety requirements of each control action (refined from Table 2), first each software safety requirement should be formalized into a formal specification such as LTL or CTL to be able to verify them against the safe behavior model of the software controller during the verification phase. For example, the refined software safety requirement SSR1.3 can be expressed as the LTL formula:

*G ((currentSpeed < desiredSpeed && distance > safeDistance && ACCMode == cruise && Brakestatus == Notapplied) $\rightarrow$ accelerateSpeed).*

This formula means that the ACC software controller must always provide an acceleration signal when the current speed of the vehicle is less than the desired speed, there is no vehicle in the lane (*distance > safe distance*), and the brake pedal is not pressed when the ACC system is in cruise mode. Second, the safe behavior model needs to be transformed into a model of the input language of a model checker such as SMV (Symbolic Model Verifier).

Once the SMV model has been constructed from the safe behavior model and the software safety requirements have been expressed in temporal logic, the formal verification activities with the SMV model checker can be performed. We ran the NuSMV 2.5.4 model checking tool on a Windows 7 PC that was equipped with an Intel Core i7-2640M CPU with 2.80 GHz, 8 GB main memory and a hard disk with a capacity of 700GB. The NuSMV tool verified the SMV model against the LTL formulae. The results of the verification indicated that all the identified STPA software safety requirements were satisfied and there is no counterexample generated because the safe behavior model itself was built using the software safety requirements identified by the STPA. To generate safety-based test cases, we converted manually the safe behavior model to the finite state machine. We used the finite state machine of the safe behavior model as input to the model-based testing tool such as ModelJUnit [15]. We created a test code generator in Java which transforms the finite state machine of the safe behavior model into ModelJUnit Java class. We selected the all-transactions, action and state test coverages. We generated automatically 487 test cases which covered the safe behavior of the ACC software controller with the action coverage 15/18, state coverage 6/6, transition coverage 15/15 and transition pair coverage 36/36. We created Java code to remove the duplicated test cases from the output of ModelJunit. The total number of final test cases was 197. We also created Java code to generate the traceability matrix between the generated test cases and the STPA-generated software safety requirements and saved the matrix into an excel sheet. Based on the traceability matrix, we calculated the coverage of the STPA-generated software safety requirements by counting the total number of STPA-generated safety requirements covered into test cases. As a result, we obtained a 100% test coverage for the STPA requirements. We also calculated the average of each STPA-generated safety requirement and control action in the safe behavior model. For example, the average of the software safety requirement SSR1.3 was 17 out of the total number of the generated test sets 197. The average of *accelerate* control action was 21 out of 197 test cases.

## 6. Conclusions

In this paper we proposed a safety engineering approach based on STPA to develop safe software. The approach can be integrated into a software development process. This approach allows the software and safety engineers to work together during the software development for safety-critical systems. Also, it highlights the advantages of applying STPA to software at the system level to identify potentially unsafe control actions of software and to derive the corresponding safety requirements that prevent software to transition into a hazardous state. The proposed approach is general and can be applied to any software; however, because the software development process as described in the safety standard ISO26262 is subdivided into sub-phases according to the V-Model, we believe that our approach can especially be adapted to be used in the context of this standard as a means to support the development of safe automotive software or to evaluate existing automotive safety-critical software.

The limitations of the proposed approach are that the main steps require manual interventions performed by the safety analyst and the difficulty of using formal verification in practice. Therefore, we plan to develop tool support that will implement the main activities of the safety verification and testing in the steps 2–4 and will be integrated into our extensible XSTAMPP safety engineering platform that was used in the main activities of step 1. This tool will help the software and safety engineers during the development process of safe software. To evaluate the proposed approach, we selected the safety-critical software of an Adaptive Cruise Control System (ACC) to explore the application of the proposed safety engineering approach to a real industrial system in the automotive domain. In this direction, we plan to conduct two more case studies. The first case study will be conducted during the development of a simulator of ACC with LEGO-mindstorm robots following the V-Model software development process. A software developer and a safety analyst will work together to develop safe software for the ACC based on the proposed approach. The second case study will be performed based on the existing software of an ACC with an industrial partner to explore the application of our approach to existing software. We also plan to evaluate the proposed approach during the development of an agile software development process.